Simulating lattice thermal conductivity in semiconducting materials using high-dimensional neural network potential

Emi Minamitani[1,2,3], Masayoshi Ogura[2], Satoshi Watanabe[2,4]

[1]*Institute for Molecular Science, Japan.*

[2]*Department of Materials Engineering, The University of Tokyo, Japan.*

[3]*Japan Science and Technology Agency, PRESTO, Japan.*

[4]*Center for Materials Research by Information Integration, Research and Services Division of Materials Data and Integrated System, National Institute for Materials Science, Japan.*

Abstract

We demonstrate that a high-dimensional neural network potential (HDNNP) can predict the lattice thermal conductivity of semiconducting materials with an accuracy comparable to that of density functional theory (DFT) calculation. After a training procedure based on the force, the root mean square error between the forces predicted by the HDNNP and DFT is less than 40 meV/Å. As typical examples, we present the results for Si and GaN bulk crystals. The deviation from the thermal conductivity calculated using DFT is within 1% at 200 to 500 K for Si and within 5.4% at 200 to 1000 K for GaN.

Heat generation in semiconducting materials has become a critical problem in modern nanoscale electronics. As the electric device size decreases, the power density and device temperature increase. This becomes one of the major factors contributing to degradation of the device performance and reliability[1]. To design semiconductor materials with better thermal manageability, efficient methods for theoretical simulation of the thermal conductivity are highly demanded.

The main carrier of heat in semiconductors is the phonon, which is the quantum of lattice vibration. Current methods of simulating the lattice thermal conductivity can be classified into three categories[2,3]: 1) anharmonic lattice dynamics (ALD) in combination with phonon transport calculation using the Boltzmann transport equation (BTE) and Fourier's law[4–8], 2) equilibrium molecular dynamics (EMD) using the Green–Kubo formula[9–11], and 3) direct evaluation of the heat flux by nonequilibrium molecular dynamics (NEMD)[12,13]. In these theoretical frameworks, accurate prediction of the interatomic force in the solid is essential. Density functional theory (DFT) calculation is one of the most well-established techniques for accurate force prediction, including the effect of the electronic state change with atomic displacement. However, the high computational cost limits the application of DFT calculation in thermal conductivity simulations. Although a combination of DFT calculations with the ALD[7] or EMD[14] approach has been successful in accurate prediction of the lattice thermal conductivities of semiconductor crystals, the application of this technique to systems with more complex structures such as defective or disordered ones is not realistic because of the rapid increase in computational cost with increasing system size. Regarding NEMD, direct combination with DFT calculation is almost impossible, because the system size used in

NEMD simulations must be much larger to reproduce a reasonable temperature gradient[13]. The computational time can be reduced by using the empirical potential, but the accuracy is insufficient compared to that of DFT-based calculation[8]. An alternative simulation technique that can resolve the trade-off between the accuracy of the force prediction and the computational cost is urgently needed.

Machine learning techniques are a promising approach[15,16], and applications to solid-state physics have been rapidly developed in recent years. As shown in early works by Behler et al.[17,18] and subsequent studies[19–25], the high-dimensional neural network potential (HDNNP) can describe the relationship between the total energy of a system and its atomic arrangement. It is naturally expected that the force acting on atoms can also be described by the HDNNP, because the derivative of the total energy with respect to the atomic displacement gives the force. Force prediction in semiconducting materials using an HDNNP or other machine learning techniques has already been reported in several studies[26,27]. However, the accuracy of the prediction is limited to on the order of 100 meV/Å, which is much larger than the DFT accuracy (a few tens of meV/Å) required to simulate the thermal conductivity. Here, we show that much higher accuracy can be obtained by training HDNNP parameters with a focus on force fitting. We chose crystalline Si and GaN as representative semiconducting materials with one and two atom types, respectively. In both systems, we obtained an HDNNP that can predict the force with the accuracy of DFT. The obtained phonon frequency and thermal conductivity are within 5.4% of those calculated by DFT.

In this study, we adopted the HDNNP model developed by Behler et al[16,17]. In this model, the total energy of the system ($E_{\text{tot}}$) is expressed as the sum of the energy contributions from each atom ($E_i$), i.e., $E_{tot} = \sum_i E_i$. Here we neglect the effect of the long-range electrostatic potential, and $E_i$ is determined by the local atomic environment, which is described by the cutoff and symmetry functions. The cutoff function determines the sphere of the local environment, and the symmetry functions represent the radial and angular distributions of neighboring atoms. Here, we use the following cutoff function:

$$f_c(R_{ij}) = \begin{cases} \tanh^3\left(1 - \dfrac{R_{ij}}{R_c}\right) & R_{ij} \leq R_c \\ 0 & R_{ij} > R_c \end{cases}, \quad (1)$$

where $R_c$ is the cutoff distance, and $R_{ij}$ is the interatomic distance. For the symmetry functions, we use the following three types of functional forms:

$$G_i^1 = \sum_{j=1}^{N_{atom}} f_c(R_{ij}), \quad (2)$$

$$G_i^2 = \sum_{j=1}^{N_{atom}} e^{-\eta(R_{ij} - R_s)^2} f_c(R_{ij}), \quad (3)$$

$$G_i^4 = 2^{1-\zeta} \sum_{j \neq i}^{N_{atom}} \sum_{k \neq i,j}^{N_{atom}} \left[(1 + \lambda \cos\theta_{ijk})^\zeta e^{-\eta(R_{ij}^2 + R_{ik}^2)} f_c(R_{ij}) f_c(R_{ik})\right]. \quad (4)$$

Here, $N_{atom}$ represents the number of atoms inside the cutoff sphere. The hyperparameters $R_c, R_s, \lambda, \zeta,$ and $\eta$, which should be set before the HDNNP is trained, must be tuned for better prediction performance. The values of the hyperparameters used in this work are summarized in the supplementary materials.

As a simple example, let us consider a neural network (NN) consisting of a single hidden layer with $N_n$ nodes and $N_s$ input nodes associated with the symmetry functions. The atomic energy and the force with respect to the atomic displacement along the Cartesian coordinate $R_i^\nu$ ($\nu = x, y, z$) output by the NN ($E_i$ and $F_i^\nu$, respectively) are given by the

expressions

$$E_i = f_a^2 \left[ w_{01}^2 + \sum_{j=1}^{N_n} w_{j1}^2 f_a^1 \left( w_{0j}^1 + \sum_{\mu=1}^{N_s} w_{\mu j}^1 G_i^\mu \right) \right], \quad (5)$$

$$F_i^\nu = -\frac{\partial E_{tot}}{\partial R_i^\nu} = -\sum_{j=1}^{N} \frac{\partial E_j}{\partial R_i^\nu} = -\sum_{j=1}^{N} \sum_{\mu=1}^{N_s} \frac{\partial G_j^\mu}{\partial R_i^\nu} \frac{\partial E_j}{\partial G_j^\mu}, \quad (6)$$

where $w_{0j}^k$, $w_{\mu \neq 0\ j}^k$, and $f_a^k$ are the bias, weight, and activation functions in the $k$-th layer, respectively. $N$ represents the total number of atoms in the system. These formulas for the atomic energy and force can be extended to the deeper NN and subnet structure in the HDNNP.

The above HDNNP model was implemented in our homemade code, which is publicly accessible via a Github repository[28]. Training of the weights and biases in the NN by back-propagation and the differentiation required for the force calculation were implemented using the Chainer[29,30] library. The loss function for the training procedure is defined as

$$L = \alpha \times \text{RMSE}(\{F_i^\nu\}) + (1 - \alpha) \times \text{RMSE}(\{E_i\}), \quad (7)$$

where RMSE({}) is the sum of the root mean square error between the HDNNP prediction and DFT training data. Note that the units of the RMSE of the energy and force are eV/atom and meV/Å, respectively. Here, the loss function $L$ is evaluated using the unitless RMSE values.

The training data sets for Si and GaN were generated from a combination of a classical molecular dynamics (MD) simulation using LAMMPS[31,32] and DFT calculation using the Vienna Ab initio Simulation Package (VASP)[33–36]. A crucial point for good training of an HDNNP is the randomness of the atomic configurations in the training data, which requires a long MD simulation and sparse sampling of the MD trajectories. Because of this sparseness, it is more efficient to generate structures using classical MD with a well-established interatomic potential and then evaluate the energies and forces of structures sampled from the MD trajectories by DFT calculation than to perform the entire procedure using DFT. We first conducted a classical MD simulation using the Stillinger–Weber potential[37] using LAMMPS for 10,000 steps at 0.001 ps time intervals at several temperatures. Here we focus on the thermal properties; thus, we need an HDNNP that expresses well the forces acting on atoms in the atomic arrangements that may appear during thermal vibration. The data for configurations with a large atomic displacement and/or bond breaking, which appear at high temperatures, are redundant for this purpose. Therefore, we selected temperatures between room temperature and the melting temperature for the classical MD simulation. For Si, we choose 300, 500, 700, and 900 K. For GaN, we choose 300, 700, 1100, 1500, 1900, and 2300 K.

Then, we randomly selected 100 snapshots from each MD trajectory and performed DFT calculations for the corresponding atomic configurations without structure relaxation by VASP with the projector augmented wave potential[38]. LDA and GGA-PBE[39] exchange-correlation functional were used for Si and GaN. The cutoff energy of the plane wave basis was set to 550 eV. In all these MD and DFT calculations, we chose unit cell sizes of 64 atoms for Si and 32 atoms for GaN. To increase the variation in the atomic configurations in the training data and extend the flexibility of the HDNNP[40], we also included the data obtained using slightly different lattice constants. The total numbers of the atomic configurations in the DFT data sets were 2800 and 3000 for Si and GaN, respectively.

In the training procedure using the above data set, we used an NN topology with two hidden layers. The number of nodes in each layer was set to 500, and the hyperbolic tangent was adopted as the activation function. Adam (Adaptive Moment Estimation) was chosen as the optimization algorithm[41].

In Fig. 1, we compare the forces in the Si and GaN systems predicted by DFT and the HDNNP. During training, 90% of the DFT data sets were used as training data, and the remaining 10% were used as validation data. We set the parameter $\alpha = 0.99$ in the loss function in Eq. (7) so that the RMSE of the force is dominant in the training procedure. The final RMSE of the force prediction in the validation data is 25.5 meV/Å for Si and 37.8 meV/Å for GaN. Even at the very low weight of the loss function, the final RMSE of the total energy prediction for the validation data reaches 32.7 meV/atom for Si and 66.5 meV/atom for GaN (see the supplementary materials). We found that the lower value of $\alpha$ increases the RMSE of the force prediction, which indicates that focusing on the force itself during training is important to obtain more accurate force prediction.

Next, to check that the force accuracy is sufficient for simulation of the phonon-related thermal properties, we evaluated the phonon dispersions in Si and GaN crystals by a combination of the HDNNP and phonopy[42]. We used HDNNP to predict the forces for the irreducible displacement patterns given by phonopy. For comparison, we also performed a phonon dispersion calculation using a combination of VASP and phonopy.

The phonon dispersion curves obtained using HDNNP are in good agreement with the DFT calculation results and previous reports[43,44] for both Si and GaN, as shown in Fig. 2. Note that we focus on the comparison between the HDNNP and DFT results, and we did not include the non-analytic correction for LO-TO splitting[45]. Thus, the frequencies of the optical modes in GaN differ from the experimental values.

Then, we simulated the lattice thermal conductivity based on ALD by combining the HDNNP and phono3py package[4], using a procedure similar to that for the phonon dispersion calculation. The irreducible displacements for evaluating the third-order potential for three phonon processes were obtained using phono3py, and the forces acting on atoms in each displacement pattern were predicted using the HDNNP. The phonon–phonon interaction strength and the corresponding lifetime were extracted from these force data; then, the lattice thermal conductivity ($\kappa$) was calculated using the single-mode relaxation time approximation of the linearized phonon BTE. Figure 3 compares the temperature dependence of the thermal conductivity obtained from the force predictions of the HDNNP and VASP calculations. We note that the thermal conductivity in GaN is underestimated compared to a previous report[46]. The reason might be the small cell size in the thermal conductivity simulation (32 atoms in this study and 108 atoms in Ref. 46). Here we focus on the reproducibility of the DFT calculation by the HDNNP; thus, we do not discuss this point in detail. The calculation results from the HDNNP and DFT calculation under the same simulation conditions for both Si and GaN are in good agreement, indicating the strong potential of the HDNNP for application in thermal conductivity simulations. The deviation from the DFT calculation results is within 1% at 200 to 500 K for Si and within 5.4% from 200 to 1000 K for GaN. For example, in Si, the $\kappa$ value at 300 K is estimated as 110.4 $W/m \cdot K$ from the HDNNP results and 112.1 $W/m \cdot K$ from the DFT results. In GaN, the thermal conductivity along the in-plane direction, $\kappa_\parallel$, and along the out-of-plane direction, $\kappa_\perp$, at 300 K are 275.5 and 309.7 $W/m \cdot K$, respectively. These values are in good agreement with the DFT calculation results, $\kappa_\parallel = 274.2$ and $\kappa_\perp = 325.5$ $W/m \cdot K$, respectively.

Finally, we comment on the computational time required to calculate the thermal conductivity using DFT and the HDNNP. For Si, the DFT calculation for a total of 111 displacement patterns required 24 CPUs and 4.5 h, and the time was reduced to 1 CPU and 10 s for the HDNNP calculation. For GaN, the DFT calculations for a total of 582 displacements required 24 CPUs and 21.0 h, and the time was reduced to 24 CPUs and 1.5 min. Including the time required to prepare the data sets for training, the total computational time for the HDNNP may be longer than that for direct DFT calculation of the thermal conductivity based on ALD in simple bulk crystals. However, if an HDNNP can be combined with an MD simulation of the thermal conductivity and/or applied to systems with defects or random alloys such as $Si_xGe_{1-x}$, the efficiency of the HDNNP would provide a new way to investigate the thermal properties with DFT accuracy. Examination of the flexibility and extensibility of the HDNNP in these directions remains as future work.

References

[1] P. Ball, Nature **492**, 174 (2012).

[2] D.G. Cahill, P.V. Braun, G. Chen, D.R. Clarke, S. Fan, K.E. Goodson, P. Keblinski, W.P.

King, G.D. Mahan, A. Majumdar, H.J. Maris, S.R. Phillpot, E. Pop, and L. Shi, Appl. Phys. Rev. **1**, 011305 (2014).

[3] P.K. Schelling, S.R. Phillpot, and P. Keblinski, Phys. Rev. B **65**, 144306 (2002).

[4] A. Togo, L. Chaput, and I. Tanaka, Phys. Rev. B **91**, 094306 (2015).

[5] K. Mizokami, A. Togo, and I. Tanaka, Phys. Rev. B **97**, 224306 (2018).

[6] J. Zhou, B. Liao, and G. Chen, Semicond. Sci. Technol. **31**, 043001 (2016).

[7] L. Lindsay, D.A. Broido, and T.L. Reinecke, Phys. Rev. B **87**, 165201 (2013).

[8] D.A. Broido, M. Malorny, G. Birner, N. Mingo, and D.A. Stewart, Appl. Phys. Lett. **91**, 11305 (2007).

[9] R. Kubo, M. Yokota, and S. Nakajima, J. Phys. Soc. Japan **12**, 1203 (1957).

[10] S.G. Volz and G. Chen, Phys. Rev. B **61**, 2651 (2000).

[11] A.J.C. Ladd, B. Moran, and W.G. Hoover, Phys. Rev. B **34**, 5058 (1986).

[12] F. Müller-Plathe, J. Chem. Phys. **106**, 6082 (1997).

[13] M.S. El-Genk, K. Talaat, and B.J. Cowen, J. Appl. Phys. **123**, 205104 (2018).

[14] C. Carbogno, R. Ramprasad, and M. Scheffler, Phys. Rev. Lett. **118**, 175901 (2017).

[15] L. Zdeborová, Nat. Phys. **13**, 420 (2017).

[16] J. Behler, J. Chem. Phys. **145**, 170901 (2016).

[17] J. Behler and M. Parrinello, Phys. Rev. Lett. **98**, 146401 (2007).

[18] J. Behler, Angew. Chem., Int. Ed. **56**, 12828 (2017).

[19] W. Li, Y. Ando, E. Minamitani, and S. Watanabe, J. Chem. Phys. **147**, 214106 (2017).

[20] W. Li, Y. Ando, and S. Watanabe, J. Phys. Soc. Japan **86**, 104004 (2017).

[21] N. Artrith, A. Urban, and G. Ceder, Phys. Rev. B **96**, 14112 (2017).

[22] N. Artrith and A. Urban, Comput. Mater. Sci. **114**, 135 (2016).

[23] B. Onat, E.D. Cubuk, B.D. Malone, and E. Kaxiras, Phys. Rev. B **97**, 94106 (2018).

[24] R. Kobayashi, D. Giofré, T. Junge, M. Ceriotti, and W.A. Curtin, Phys. Rev. Mater. **1**, 53604 (2017).

[25] W. Jeong, K. Lee, D. Yoo, D. Lee, and S. Han, J. Phys. Chem. C **122**, 22790 (2018).

[26] Y. Huang, J. Kang, W.A. Goddard, and L.-W. Wang, Phys. Rev. B **99**, 064103 (2019).

[27] X. Qian and R. Yang, Phys. Rev. B **98**, 224108 (2018).

[28] Our HDNNP code can be downloaded at https://github.com/ogura-edu/HDNNP.

[29] S. Tokui, K. Oono, S. Hido, and J. Clayton, "*Chainer: A Next-Generation Open Source Framework for Deep Learning,*" http://learningsys.org/papers/LearningSys_2015_paper_33.pdf.

[30] T. Akiba, K. Fukuda, and S. Suzuki, "*ChainerMN: Scalable Distributed Deep Learning Framework,*" http://learningsys.org/papers/LearningSys_2015_paper_33.pdf.

[31] S. Plimpton, J. Comput. Phys. **117**, 1 (1995).

[32] LAMMPS is provided at https://lammps.sandia.gov/.

[33] G. Kresse and D. Joubert, Phys. Rev. B **59**, 1758 (1999).

[34] G. Kresse and J. Furthmüller, Comput. Mater. Sci. **6**, 15 (1996).

[35] G. Kresse and J. Hafner, Phys. Rev. B **47**, 558 (1993).

[36] G. Kresse and J. Furthmüller, Phys. Rev. B **54**, 11169 (1996).

[37] F.H. Stillinger and T.A. Weber, Phys. Rev. B **31**, 5262 (1985).

[38] P.E. Blöchl, Phys. Rev. B **50**, 17953 (1994).

[39] J.P. Perdew, K. Burke, and M. Ernzerhof, Phys. Rev. Lett. **77,** 3865 (1996).

[40] D. Yoo, K. Lee, W. Jeong, S. Watanabe, and S. Han, arXiv:1903.04366 (2019).

[41] D.P. Kingma and J. Ba, *Adam: A Method for Stochastic Optimization,* arXiv:1412.6980 (2014).

[42] A. Togo and I. Tanaka, Scr. Mater. **108**, 1 (2015).

[43] T. Ruf, J. Serrano, M. Cardona, P. Pavone, M. Pabst, M. Krisch, M. D'Astuto, T. Suski, I. Grzegory, and M. Leszczynski, Phys. Rev. Lett. **86**, 906 (2001).

[44] A. Ward and D.A. Broido, Phys. Rev. B **81**, 085205 (2010).

[45] S. Baroni, S. De Gironcoli, A. Dal Corso, and P. Giannozzi, Rev. Mod. Phys. **73**, 515 (2001).

[46] L. Lindsay, D.A. Broido, and T.L. Reinecke, Phys. Rev. Lett. **109**, 095901 (2012).

Acknowledgments

The following financial support is acknowledged: JST PRESTO JPMJPR17I7 (E. M.), JST "Materials research by Information Integration" Initiative (MI2I) project of the Support Program for Starting Up Innovation Hub (S. W.), and MEXT KAKENHI 17H05330 (S. W. and E. M.). We thank Prof. Togo for fruitful discussions on the connection between the HDNNP and phonopy/phono3py. The calculations were performed using the computer facilities of the Institute of Solid State Physics, Information Technology Center, the University of Tokyo, and RIKEN (HOKUSAI GreatWave).

Figure captions

Figure 1

Comparison of interatomic forces in (a) Si and (b) GaN bulk crystals obtained by HDNNP and DFT calculations.

Figure 2

Comparison of phonon dispersions in (a) Si and (b) GaN obtained by HDNNP and DFT calculations. The supercell sizes for the calculations of the second-order force constant were set to 2 × 2 × 2 for the conventional unit cell in Si and 3 × 3 × 2 for the primitive unit cell in GaN. The experimental data for comparison were taken from Ref. 44 for Si and Ref. 43 for GaN.

Figure 3

Comparison of thermal conductivity in (a) Si and (b) that along the in-plane (100) direction in GaN, and (c) that along the out-of-plane (001) direction in GaN obtained by HDNNP and DFT calculations. The supercell sizes for the calculations of the third-order force constant were set to 2 × 2 × 2 for the conventional unit cell in Si and 2 × 2 × 2 for the primitive unit cell in GaN. In the calculation of the linearized Boltzmann equation with the relaxation time approximation, the Brillouin zone was sampled on an 11 × 11 × 11 mesh in all cases.

Figures

Figure 1

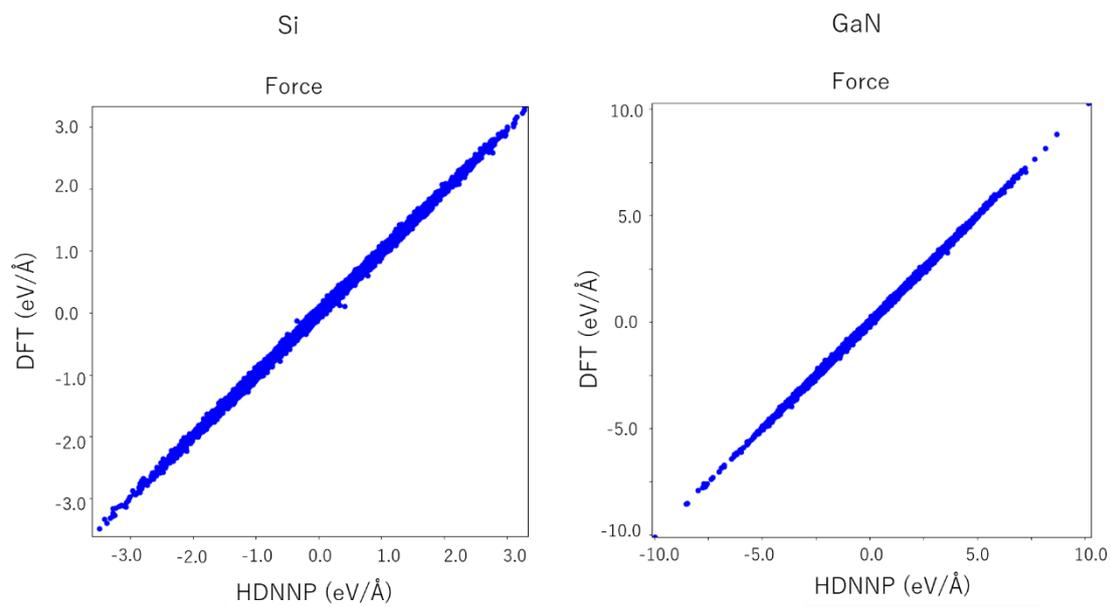

Figure 2

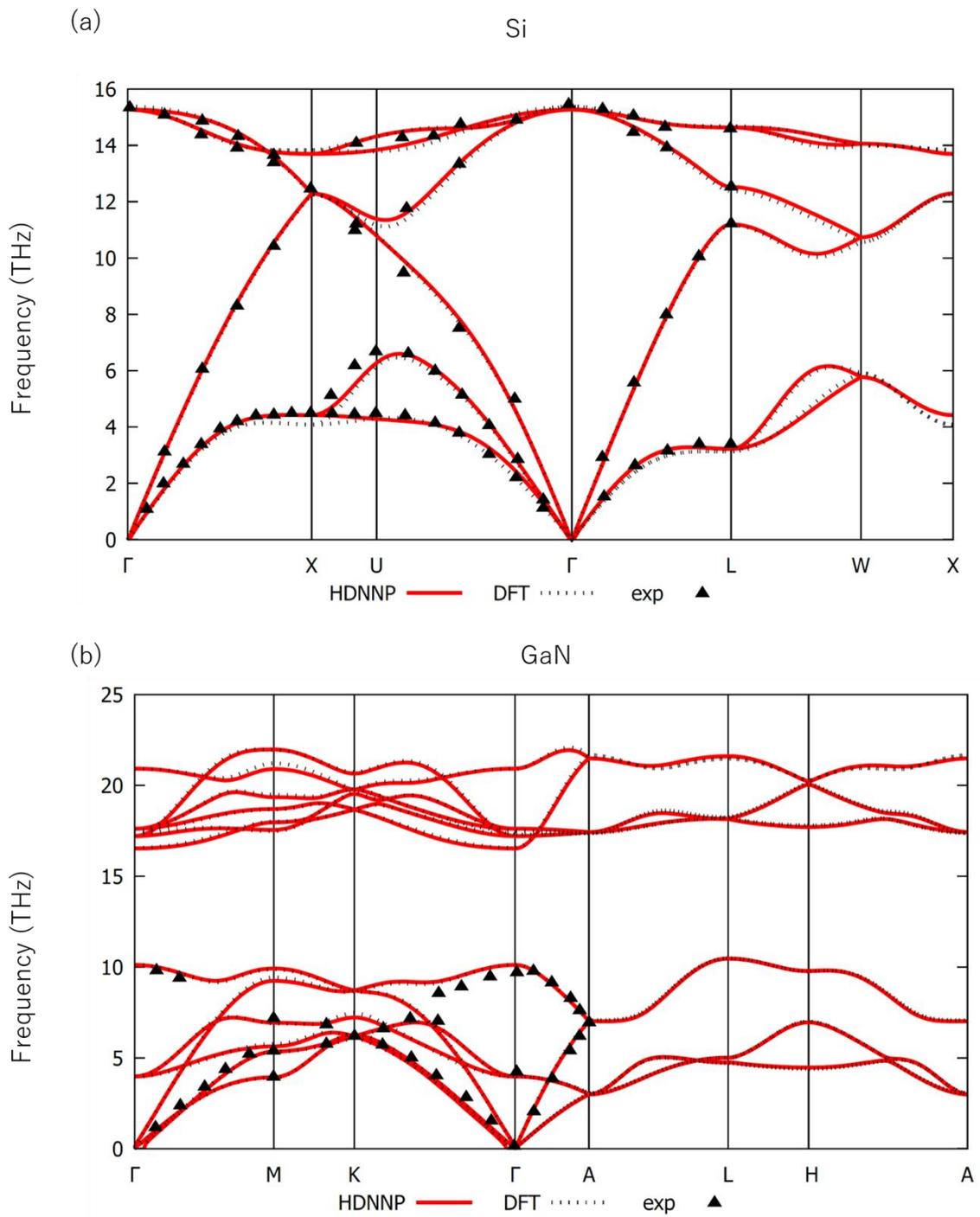

Figure 3

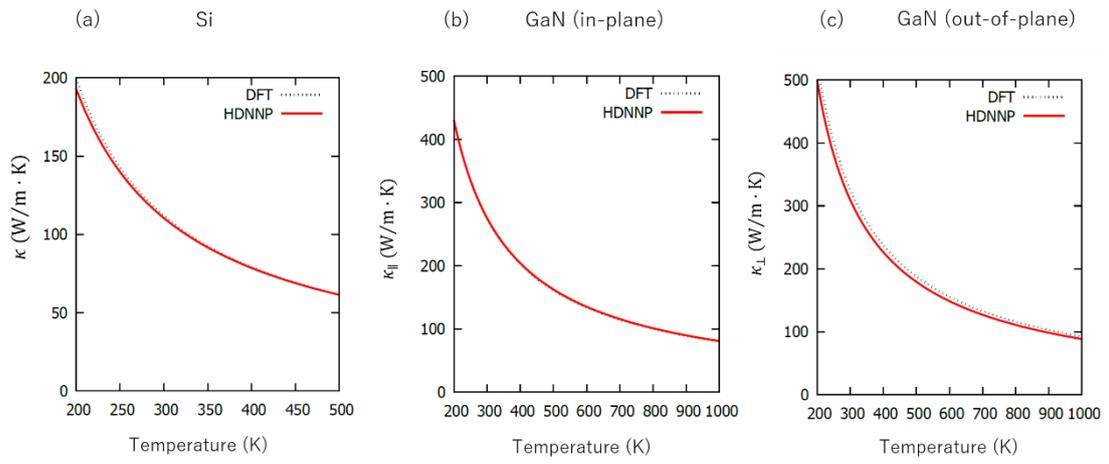

Supplementary materials for
"Simulating lattice thermal conductivity in semiconducting materials using high-dimensional neural network potential"

Emi Minamitani[1,2,3], Masayoshi Ogura[2], Satoshi Watanabe[2,4]

[1]*Institute for Molecular Science, Japan.*
[2]*Department of Materials Engineering, The University of Tokyo, Japan.*
[3]*Japan Science and Technology Agency, PRESTO, Japan.*
[4]*Center for Materials Research by Information Integration, Research and Services Division of Materials Data and Integrated System, National Institute for Materials Science, Japan.*

1. Values of hyperparameters used in this work

   The hyperparameters of radial and angular symmetry functions used in Si and GaN simulations are listed in Table S1 and Table S2, respectively.

   Table S1 hyperparameters used in Si simulation

   $G^1$ symmetry functions

   | No. | $R_c$(Å) |
   |-----|----------|
   | 1   | 6.0      |

   $G^2$ symmetry functions

   | No. | $R_c$(Å) | $\eta$ | $R_s$(Å) |
   |-----|----------|--------|----------|
   | 2   | 6.0      | 0.1    | 2.2      |
   | 3   | 6.0      | 0.1    | 2.4      |
   | 4   | 6.0      | 0.1    | 2.6      |
   | 5   | 6.0      | 0.1    | 3.0      |
   | 6   | 6.0      | 0.1    | 3.4      |
   | 7   | 6.0      | 0.1    | 3.6      |
   | 8   | 6.0      | 0.1    | 4.0      |
   | 9   | 6.0      | 0.1    | 4.4      |
   | 10  | 6.0      | 0.5    | 2.2      |
   | 11  | 6.0      | 0.5    | 2.4      |

| | | | |
|---|---|---|---|
| 12 | 6.0 | 0.5 | 2.6 |
| 13 | 6.0 | 0.5 | 3.0 |
| 14 | 6.0 | 0.5 | 3.4 |
| 15 | 6.0 | 0.5 | 3.6 |
| 16 | 6.0 | 0.5 | 4.0 |
| 17 | 6.0 | 0.5 | 4.4 |
| 18 | 6.0 | 1.0 | 2.2 |
| 19 | 6.0 | 1.0 | 2.4 |
| 20 | 6.0 | 1.0 | 2.6 |
| 21 | 6.0 | 1.0 | 3.0 |
| 22 | 6.0 | 1.0 | 3.4 |
| 23 | 6.0 | 1.0 | 3.6 |
| 24 | 6.0 | 1.0 | 4.0 |
| 25 | 6.0 | 1.0 | 4.4 |

$G^4$ symmetry functions

| No. | $R_c$(Å) | $\eta$ | $\lambda$ | $\zeta$ |
|---|---|---|---|---|
| 26 | 6.0 | 0.0 | 1 | 1 |
| 27 | 6.0 | 0.0 | 1 | 2 |
| 28 | 6.0 | 0.0 | 1 | 4 |
| 29 | 6.0 | 0.0 | 1 | 16 |
| 30 | 6.0 | 0.0 | -1 | 1 |
| 31 | 6.0 | 0.0 | -1 | 2 |
| 32 | 6.0 | 0.0 | -1 | 4 |
| 33 | 6.0 | 0.0 | -1 | 16 |
| 34 | 6.0 | 0.1 | 1 | 1 |
| 35 | 6.0 | 0.1 | 1 | 2 |
| 36 | 6.0 | 0.1 | 1 | 4 |
| 37 | 6.0 | 0.1 | 1 | 16 |
| 38 | 6.0 | 0.1 | -1 | 1 |
| 39 | 6.0 | 0.1 | -1 | 2 |
| 40 | 6.0 | 0.1 | -1 | 4 |
| 41 | 6.0 | 0.1 | -1 | 16 |

Table S2 hyperparameters used in GaN simulation

$G^1$ symmetry functions

| No. | $R_c$(Å) |
|---|---|
| 1 | 5.0 |

$G^2$ symmetry functions

| No. | $R_c$(Å) | η | $R_s$(Å) |
|---|---|---|---|
| 2 | 5.0 | 0.1 | 2.0 |
| 3 | 5.0 | 0.1 | 3.2 |
| 4 | 5.0 | 0.1 | 3.8 |
| 5 | 5.0 | 0.5 | 2.0 |
| 6 | 5.0 | 0.5 | 3.2 |
| 7 | 5.0 | 0.5 | 3.8 |
| 8 | 5.0 | 1.0 | 2.0 |
| 9 | 5.0 | 1.0 | 3.2 |
| 10 | 5.0 | 1.0 | 3.8 |

$G^4$ symmetry functions

| No. | $R_c$(Å) | η | λ | ζ |
|---|---|---|---|---|
| 11 | 5.0 | 0.0 | -1 | 1 |
| 12 | 5.0 | 0.0 | -1 | 2 |
| 13 | 5.0 | 0.0 | -1 | 4 |
| 14 | 5.0 | 0.0 | -1 | 16 |
| 15 | 5.0 | 0.0 | 1 | 1 |
| 16 | 5.0 | 0.0 | 1 | 2 |
| 17 | 5.0 | 0.0 | 1 | 4 |
| 18 | 5.0 | 0.0 | 1 | 16 |

2. Learning curves for validation data

The learning curves for the validation data as functions of number of epoch obtained in the training process of HDNNP for Si and GaN are shown in Figures S1 and S2, respectively.

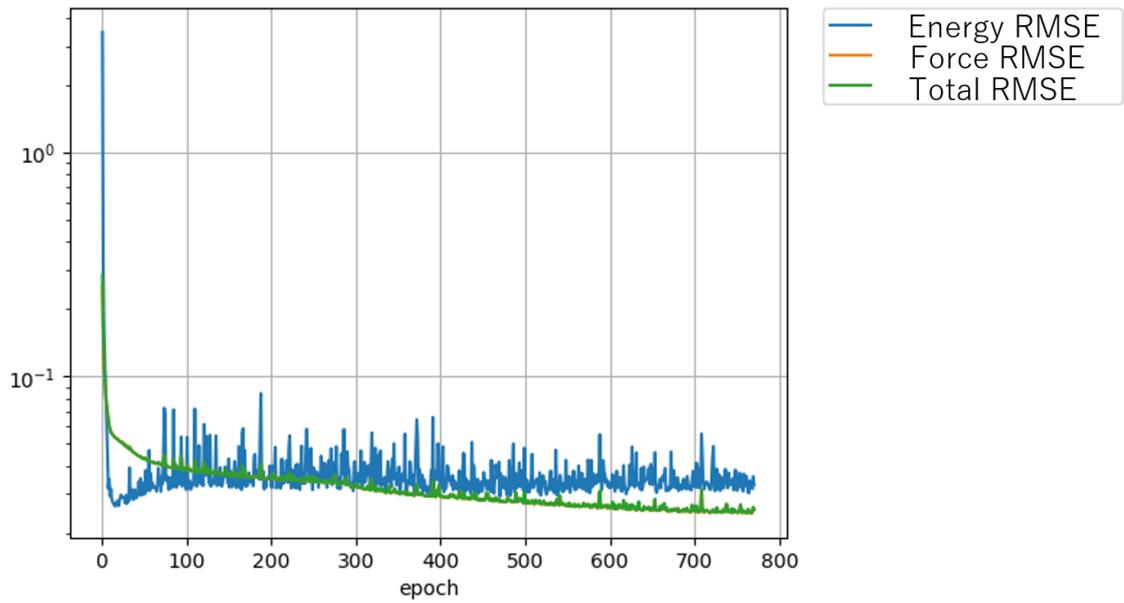

Figure S1 Learning curves of energy, force, and total RMSEs in the training procedure for Si. The units of RMSE are eV/atom for the energy and eV/ Å for the force. The total RMSE is the sum of the unitless RMSE values of the energy and force. The curves for force and total are overlap because of the large $\alpha$ in Eq.(7)

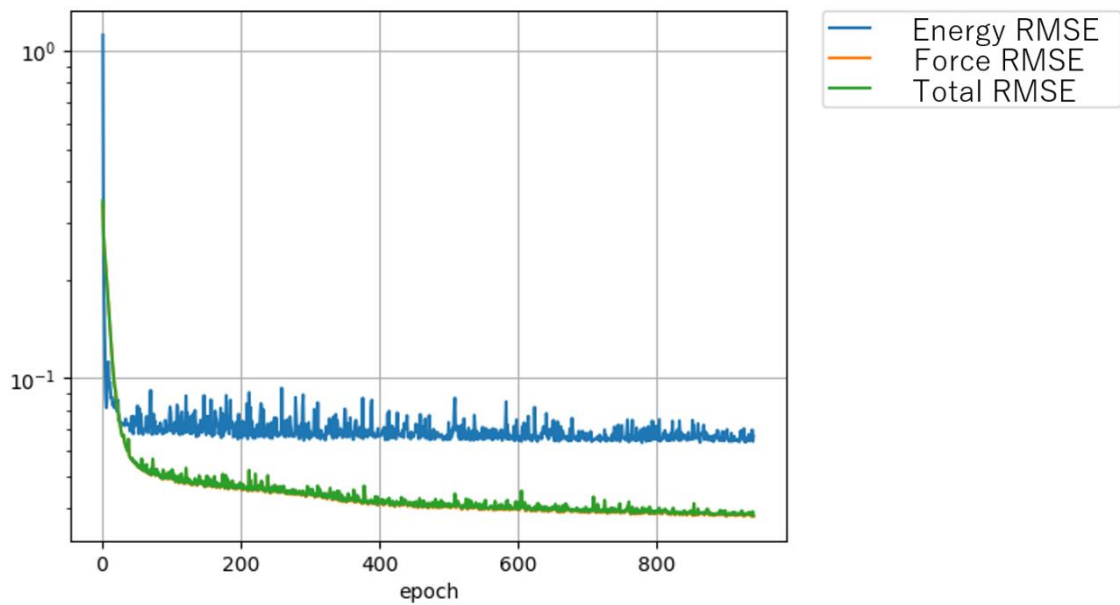

Figure S2 Learning curves of energy, force, and total RMSE in the training procedure for GaN. The units of RMSE are eV/atom for the energy and eV/ Å for the force. The total RMSE is the sum of the unitless RMSE values of the energy and force. The curves for force and total are overlap because of the large $\alpha$ in Eq.(7)